# Monte Carlo Simulation of Spin-Polarized Transport

Min Shen,  Semion Saikin,[*]  Ming-C. Cheng  and  Vladimir Privman

Center for Quantum Device Technology, Clarkson University,
Potsdam, New York 13699-5720, USA

[*]Department of Physics, Kazan State University,
Kazan, Russian Federation

**Abstract.** Monte Carlo simulations are performed to study the in-plane transport of spin-polarized electrons in III-V semiconductor quantum wells. The density matrix description of the spin polarization is incorporated in the simulation algorithm. The spin-orbit interaction terms generate coherent evolution of the electron spin polarization and also cause dephasing. The spatial motion of the electrons is treated semiclassically. Three different scattering mechanisms—optical phonons, acoustic phonons and ionized impurities—are considered. The electric field is calculated self-consistently from the charge distribution. The Monte Carlo scheme is described, and simulation results are reported for temperatures in the range 77-300 K.

## 1. Introduction.

Monte Carlo device simulation is a widely used method for modeling charge carrier transport in semiconductor devices. The approach can easily accommodate different properties of the electron transport. It is particularly well suited for highlighting the leading physical mechanisms [1]. It yields an accurate description of the device, which is not limited by the assumptions made in deriving the alternative drift-diffusion and hydrodynamic models [2,3]. Furthermore, Monte Carlo simulation can provide the physical parameters required as input for drift-diffusion and hydrodynamic models.

Spin-polarized electron transport in semiconductors has attracted significant recent interest due to its promising role in novel device structures [4-7]. Many devices utilizing spin-dependent phenomena have been proposed [8-17]. The additional spin degree of freedom, which is usually ignored in charge-transport models, can be used to encode information in the spin-polarized current. Design of new spintronic devices requires control for the spin polarization in the device channel. Presently, there are numerous difficulties in accomplishing such control. Recent experimental advances [5] have allowed efficient injection of the spin-polarized current into low-dimensional semiconductor structures [18-19] and its maintenance for up to few nanoseconds at room temperature [20]. Generally, the electron spin dynamics can be controlled by external magnetic field, local magnetic fields produced by magnetic impurities and nuclei, and spin-orbit interaction. These interactions lead to coherent evolution of carrier spin polarization and also cause spin dephasing.

Different approaches have been proposed to describe the spin-polarized current in different transport regimes. Quantum-mechanical single-particle models have been utilized for ballistic spin-polarized electron transport [15,21,22]. Semiclassical drift-diffusion models have been derived based on the two-current (spin-up and spin-down) approximation [23-25] or using the full spin-polarization vector description [26]. Recently, some nonlinear corrections in spin-polarized electron transport have attracted attention [27]. Boltzmann equations for two spin-states [28] and for spin density matrix [29,30] have also been considered.

The size of the spintronic devices is limited by the spin dephasing length, which is expected to have values of order of one micron [31,32]. For such device size, the charge distribution in the channel is far from equilibrium. Therefore, the electric field must be calculated self-consistently from the charge distribution. Monte Carlo simulation provides a natural tool for such calculations [33-35]. The simulation results are promising and consistent with the existing experimental data, though details of the Monte Carlo simulation scheme have traditionally not been presented. Moreover, some simplified assumptions were included in many studies. In the work by Bournel et al. [33], the Monte Carlo simulation has been carried out for constant in-channel electric field, while in the paper by Kiselev and Kim [34], the assumption that all the carriers have the same velocity magnitude was made and anisotropic scattering effects were ignored.

This work presents selected Monte Carlo simulation results [35] for spin-polarized electron transport in a device channel modeled as a single quantum well of a III-V heterostructure. Though the model is simple and involves certain assumptions on the device structure, we expect it to apply beyond the drift-diffusion transport regime. Moreover, additional details of the structure can be easily incorporated. The spin density matrix approach is used to evaluate the spin-polarization dynamics. The Poisson equation is solved for every sampling time step to update the electric field in the device channel. Electrons injected from the source have random momentum directions and Maxwellian distribution of magnitudes, determined by the lattice temperature. Both isotropic and anisotropic scatterings are considered. We review the model, describe the Monte Carlo procedure, and report new results on the temperature dependence of the spin dephasing length.

## 2. Spin density matrix formalism and Monte Carlo simulation.

To incorporate the electron spin dynamics into a typical Monte Carlo transport simulation model, we start with the Hamiltonian of a single conduction electron with spin,

$$H(\boldsymbol{\sigma},\mathbf{k}) = H_0(\mathbf{k}) \cdot \hat{1}_s + H_s(\boldsymbol{\sigma},\mathbf{k}) \ . \tag{1}$$

We assume that the external magnetic field is zero. The operator $\hat{1}_s$ on the right-hand side of Eq. (1) is the unity operator in the spin variables; $H_0$ is the spin-independent self-consistent single-electron Hamiltonian in the Hartree approximation,

$$H_0 = -\frac{\hbar^2}{2m^*}k^2 + V_H(\mathbf{r}) + H_{\text{e-ph}} + H_{\text{ph}} + V_{\text{imp}} \ , \tag{2}$$

including also interactions with phonons and static imperfections. The term $V_{\text{imp}}$ describes ionized nonmagnetic impurities, quantum well roughness and other static imperfections of its structure. The terms labeled "e-ph" and "ph" represent the electron-phonon interaction and the phonon mode Hamiltonian, respectively. The Hartree potential $V_H$ accounts for the electron-electron interactions. It is determined by the appropriate Poisson equation [36],

$$\nabla^2 V_H = -\frac{e^2}{\varepsilon_s}\left(\sum_j |\psi_j(\mathbf{r})|^2 - N_D\right) \ , \tag{3}$$

where $\varepsilon_s$ is the material permittivity, $|\psi_j(\mathbf{r})|^2$ is the probability density to find the $j^{\text{th}}$ electron at $\mathbf{r}$, and $N_D$ is the donor concentration. The second term on the right-hand side of Eq. (1) describes the spin dependent interactions with magnetic impurities and nuclear spins, and also the spin-orbit interaction. In this work, we only consider the effects of the spin-orbit interaction, which has been identified [37] as the main cause of the spin relaxation in III-V semiconductors at high (77-300 K) temperatures.

The appropriate description of the electron spin in an open quantum system can be given by the spin density matrix [38],

$$\rho_\sigma(t) = \begin{pmatrix} \rho_{\uparrow\uparrow}(t) & \rho_{\uparrow\downarrow}(t) \\ \rho_{\downarrow\uparrow}(t) & \rho_{\downarrow\downarrow}(t) \end{pmatrix} \ , \tag{4}$$

where $\rho_{\uparrow\uparrow}$ and $\rho_{\downarrow\downarrow}$ are the probabilities to find the electron with spin up or spin down. The (complex-conjugate) matrix elements $\rho_{\uparrow\downarrow}$ and $\rho_{\downarrow\uparrow}$ describe the linear superposition of the spin-up and spin-down states. The density matrix (4) can be parameterized by the (real) electron spin-polarization vector as $S_\zeta(t) = Tr(\sigma_\zeta \rho_\sigma(t))$, where $\zeta = x, y, z$, and $\sigma_\zeta$ are the Pauli matrices [38].

To specify the spin-orbit interaction term, we consider a single III-V asymmetric quantum well grown in the (0, 0, 1) crystallographic direction. The main spin-orbit contributions in this case are due to the bulk inversion asymmetry of the crystal—the Dresselhaus mechanism [39,40],

$$H_D = \beta \langle k_z^2 \rangle (k_y \sigma_y - k_x \sigma_x) \ , \tag{5}$$

and inversion asymmetry of the quantum well—the Rashba mechanism [41],

$$H_R = \eta(k_y \sigma_x - k_x \sigma_y) \ . \tag{6}$$

To specify the momentum and spin-polarization vector components, we use the coordinate system where $x$ is the direction of the electric field along the channel, while $z$ is orthogonal to the quantum well plane. Moreover, the axes are oriented along the principal crystal axes, and the quantum well is assumed narrow, such that $k_x^2, k_y^2 \ll \langle k_z^2 \rangle$. The latter properties are important for the form assumed for the Dresselhaus spin-orbit interaction term [40]; see Eq. (5).

For submicrometer devices with smooth potential, in the considered temperature regime ($T$ = 77-300 K), the spatial electron motion can be assumed semiclassical and described by the Boltzmann equation; see [36]. The electrons travel along classical "localized" trajectories between the scattering events. The scattering rates are given by the Fermi Golden Rule, and the scattering events are instantaneous [36]. The phonon bath in Eq. (2) is assumed to remain in thermal equilibrium with the constant lattice temperature $T$. In this case, the Monte Carlo simulation approach can be applied to the spatial transport [1-3]. We assume here that the back reaction of the electron spin evolution on the spatial motion is negligible owing to the small value of the electron momentum-state splitting due to spin-orbit interaction in comparison with its average momentum. This is consistent with the original model of the D'yakonov-Perel' spin-relaxation mechanism [42].

In the simulation model, electrons propagate with constant momentum during the time $\tau$, which is the smaller of the sampling time step $\Delta t$ and the time left to the next scattering event or from scattering to the next sampling. The propagation momentum is set equal to the average value of the momentum of a particle moving with constant acceleration during this time interval. We term this motion "free flight." For each "free flight" time interval, $\tau$, the spin density matrix evolves according to

$$\rho_\sigma(t+\tau) = e^{-i(H_R+H_D)\tau/\hbar} \rho_\sigma(\tau) e^{i(H_R+H_D)\tau/\hbar} \; . \tag{7}$$

Equation (7) is equivalent to rotation of the spin polarization vector about the effective magnetic field determined by the direction of the electron momentum. We assume that there are no electron spin-flip events accompanying momentum scattering [43]. The exponential operators in Eq. (7) can be written as (2×2) scattering matrices,

$$e^{-i(H_R+H_D)\tau/\hbar} = \begin{pmatrix} \cos(|\alpha|\tau) & i\frac{\alpha}{|\alpha|}\sin(|\alpha|\tau) \\ i\frac{\alpha^*}{|\alpha|}\sin(|\alpha|\tau) & \cos(|\alpha|\tau) \end{pmatrix}, \tag{8}$$

with the Hermitean conjugate of Eq. (8) for the operator $e^{i(H_R+H_D)\tau/\hbar}$. The sampling time step $\Delta t$ should be taken short as compared to all the dynamical time scales, in a proper Monte Carlo simulation. In Eq. (8), $\alpha$ is determined by the spin-orbit interaction terms, Eqs. (3,4),

$$\alpha = \hbar^{-1}\left[\left(\eta k_y - \beta\langle k_z^2\rangle k_x\right) + i\left(\eta k_x - \beta\langle k_z^2\rangle k_y\right)\right] \; . \tag{9}$$

During the "free flight," the spin dynamics of a single electron spin is coherent; see Eq. (7). However, stochastic momentum fluctuations due to electron scattering events, produce the distribution of spin states, thus causing effective dephasing at times $t > 0$.

The spin polarization, $\langle S_\zeta(\mathbf{r},t)\rangle$, of the current can be obtained by averaging $S_\zeta$ over all the electrons in a small volume $dv$, which is located at the space position $\mathbf{r}$, at time $t$. The absolute value of the average spin polarization vector is in the range $|\langle \mathbf{S}(\mathbf{r},t)\rangle| \leq 1$. If $|\langle \mathbf{S}(\mathbf{r},t)\rangle|$ is equal to 1, the electric current is completely spin-polarized. The components $\langle S_\zeta(\mathbf{r},t)\rangle$ define the orientation of the spin polarization, and evolution of the spin polarization vector may be viewed as consisting of coherent motion (rotation) and loss of polarization (reduction of magnitude) due to electron spin dephasing [40,42].

The Monte Carlo simulation is carried out by sequentially performing free-flight and scattering calculations for all the particles. The next-scattering-event time is generated as $\delta t_{\text{scat}} = -(\ln p)/\Gamma$, where $p$ is a random number between 0 and 1, and $\Gamma$ is the total scattering rate including the self-scattering rate [2,3,44] that accounts for fictitious scattering introduced to make $\Gamma$ constant. The sampling time step $\Delta t$ is specified small enough to properly update the particle motion and the electric field. The choice of the value of $\Delta t$ is based on the stability criteria [45]. The momentum increment and the distance of the "free flight" are calculated as

$$\Delta \mathbf{k}\hbar = e\mathbf{E}\tau, \quad \Delta \mathbf{r} = \frac{\hbar\,(\mathbf{k} + \Delta\mathbf{k}/2)}{m}\tau, \tag{10}$$

where $-e$ is the electron charge and $\mathbf{E}$ is the applied electric field. Based on the above discussion, the additional calculation needed to follow the spin polarization evolution of each particle, consists of an update of the spin density matrix at the end of each "free flight" time step, by using Eqs. (7,8).

It is assumed that the electrons are confined in the 1st (lowest) subband and that their *z*-direction motion is steady-state and defined by the shape of the quantum well. In the scattering event calculations, three in-plane (*xy*) scattering mechanisms are included in the simulation: optical phonon scattering, acoustic phonon scattering (for the scattering rates, see Sect. 2.6 of [45]), and separated impurity scattering (for the scattering rate, see Sect. 7 of [46]).

The selection of the scattering mechanism is performed by defining

$$\Lambda_n(E_\mathbf{k}) = \sum_{j=1}^{n} W_j(E_\mathbf{k})/\Gamma, \quad n = 1,\ 2,\ 3, \tag{11}$$

where $W_j(E_\mathbf{k})$ is the integral scattering rate for the *j*th mechanism. The *n*th scattering mechanism is chosen if a random number $p$ falls between $\Lambda_{n-1}(E_\mathbf{k})$ and $\Lambda_n(E_\mathbf{k})$.

In the scattering calculation, the in-plane projection of the electron momentum $k' = |\mathbf{k}'|$ is obtained from the energy conservation relation as $k' = \sqrt{2mE_{k'}}/\hbar$, where $E_{k'} = E_k \pm \hbar\omega$ for the optical phonon scattering, and $E_{k'} = E_k$ for the acoustic-phonon and impurity scattering.

The following boundary conditions are assumed. Electrons are injected at the emission boundary with the kinetic energy

$$E = -k_B T \ln p \qquad (12)$$

($T$ is the lattice temperature), and the injection angle (with respect to the $x$ axis) is randomly distributed between $-\pi/2$ and $\pi/2$. The electrons that fly beyond the collection boundary (and some that return through the injection boundary) are absorbed, and a new electron is emitted whenever there is an electron absorbed. The electric potential is the solution of the Poisson equation with the boundary conditions specified by the voltage applied to the device.

## 3. Simulation results.

For simulations, we have used the structure with the 0.55 μm channel length and infinite width, Fig. 1(a). The confining potential is assumed to be a single asymmetric $In_{0.52}Al_{0.48}As/In_{0.53}Ga_{0.47}As/In_{0.52}Al_{0.48}As$ quantum well, Fig. 1(b), in the one-subband approximation. The width of quantum well is $d = 20$ nm. The structure is highly n-doped with donor concentration $N_D = 10^{12}$ cm$^{-2}$. We assume that all the donors are ionized, and the equilibrium electron concentration in the channel is equal $N_D$. The calculated energy of the 1$^{st}$ subband is $E_1 \approx 0.2$ eV. The energy splitting between the 1$^{st}$ and 2$^{nd}$, excited, subband is estimated as $\Delta E_{12} \approx 60-70$ meV. This value in turn defines the range of the drain-source voltage values, $V_{DS}$, for which the one-subband approximation model is valid. The values of the electron spin-orbit coupling constants $\eta = 0.074$ eV·Å and $\beta = 32.2$ eV·Å$^3$ were taken from [47] and [48], respectively, while other material parameters were adopted from [49].

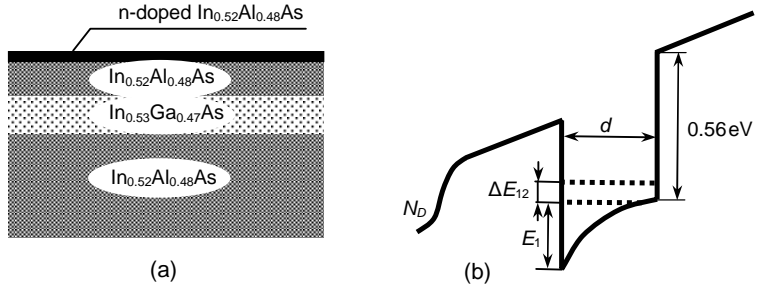

**Fig. 1.** The device structure, (a), and the confining potential, (b).

In simulation, the total number of particles in the channel was $N = 55000$, with periodic boundary conditions, and the sampling time step was $\Delta t = 1$ fsec. To achieve the steady-state transport regime, we ran the simulation program for 20000 time steps, and collected data only during the last 2000 time steps.

The simulated energy profile and in-channel electron concentration are shown on Fig. 2. In the considered range of the applied voltages, the steady-state charge distribution in the device channel is nearly constant. The injection region with varying charge distribution (up to 0.01μm) can be considered as quasi-ballistic, where electrons experience strong acceleration, Fig. 2(b), while the transport in the rest of the device is effectively drift-diffusive. The simulated steady-state distributions of the spin polarization for three different injected polarizations: along the *x*, *y*, and *z* axes, are shown in Fig. 3.

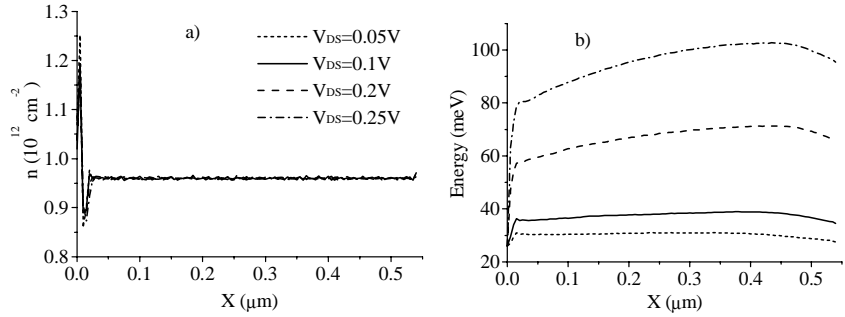

**Fig. 2.** The calculated electron transport characteristics: a) electron concentration in the channel; b) average energy profile, as functions of *x*, at $T = 300$ K and $V_{DS} = 0.05$-$0.25$ V.

In the considered model, the spin dephasing is influenced by the electron transport parameters and instantaneous orientation of the spin polarization vector. In the quasi-ballistic transport region, the spin-polarization decreases significantly, as can be seen in Fig. 3. This likely reflects the energy dependence of the scattering rates.

Due to anisotropy of the spin-orbit interaction terms, Eqs. (5,6), the spin dephasing rate is different for different orientations of the spin polarization in the drift-diffusive transport region. This leads to variations in the dephasing rate for the spin-polarized current with the injected spin polarization along the *x* and *z* directions, Fig. 3(d). For these cases, the spin polarization vector largely rotates in the *xz*-plane, Figs. 3(a) and 3(c). The dephasing will be stronger for the polarization vector oriented in the *z* direction. This can be explained by the following observation. In the considered structure, the Rashba spin-orbit coupling is considerably stronger than the Dresselhaus coupling, $\eta/(\beta\langle k_z^2\rangle) \approx 5.3$. Thus, the term proportional to $k_y$, see Eq. (6), is primarily responsible for the spin dephasing [8]. It will not affect the polarization vector oriented in the *x* direction, due to proportionality to $\sigma_x$.

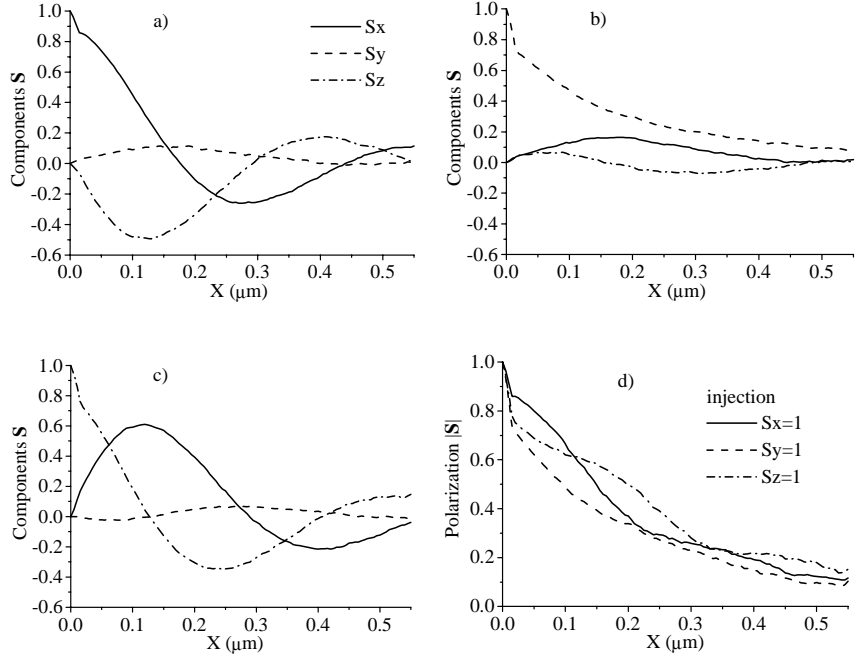

**Fig. 3.** The steady-state spin polarization, **S**, in the channel, for $V_{DS}$ = 0.1 V, $T$ = 300 K, for three different injected polarizations; a), b), c) the components of the spin polarization vector; d) the magnitude of the spin polarization vector.

The spin dephasing along the channel is not a simple exponential decay. However, we can still identify the spin dephasing length, $l_s$, as the distance over which the spin polarization is reduced by the factor of $e$ from the injected value. For higher values of the applied voltage, at low temperatures spin depolarizes faster, Fig. 4. This can be an effect of stronger scattering. However, at room temperature we observe the opposite dependence, due to larger drop of polarization in the ballistic region for smaller values of the applied voltage.

## 4. Model improvements.

Our simulation model has incorporated the leading, D'yakonov-Perel'-type spin dephasing mechanism only, which should be dominant in the semiclassical transport regime. For more accurate estimations of the electron spin dephasing, additional mechanisms should be considered [50].

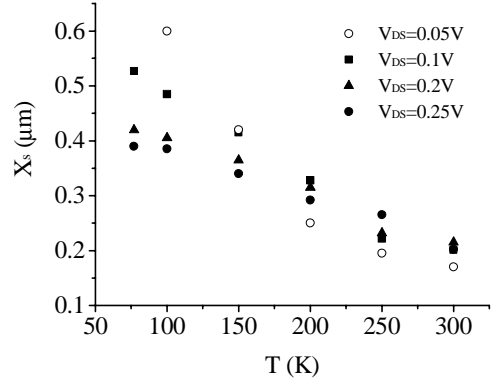

**Fig. 4.** Spin dephasing length as a function of the temperature for different values of the applied voltage (for the injected spin polarization $S_x = 1$).

In narrow band gap semiconductors such as InGaAs, the Elliott-Yafet spin-dephasing mechanism [43] can play an important role. Due to admixing of the hole states in the conduction electron wave functions, the electron spin can flip with some probability even at a non-magnetic impurity. This mechanism can be integrated in the Monte Carlo scheme in the scattering calculation, together with the momentum scattering. Another possible spin dephasing mechanism arises due to the electron-electron interaction [51]. While this does not contribute to the electron momentum and energy relaxation, the current spin polarization can be suppressed [51]. This mechanism could be important for high electron concentrations.

The validity of the one-subband approximation model is in doubt for room-temperature electron transport. In the considered case, it can be argued that the inter-subband electron scattering only contributes corrections to spin dephasing [35]. However, for more accurate calculations, inter-subband processes should be incorporated into the simulation model.

The specific device structure can also lead to additional spin dephasing mechanisms. For example, the current spin dephasing due to magnetic field created by the ferromagnetic source and drain in a spin-FET [52] may be more critical than the considered D'aykonov-Perel'-type spin relaxation.

## 5. Conclusions.

A Monte Carlo model for simulation of the spin-polarized electron transport in submicrometer device structures has been developed. The electron spin polarization is described by the spin density matrix, while spatial electron motion is treated semiclassically. The coherent dynamics of the current spin polarization, and spin dephasing, are determined by the spin-orbit interaction. The electric field in the device is evaluated self-consistently from the charge distribution. The phonon and impurity electron momentum scattering mechanisms are incorporated in the simulation. The steady state spatial distribution of the current spin-polarization vector has been simulated. The temperature dependence of the spin dephasing length was calculated for the range 77-300K. The estimated value of the spin dephasing length at room temperature is of the order of 0.2 microns.

We thank Drs. L. Fedichkin, A. Shik and I. D. Vagner for helpful discussions. This research was supported by the National Security Agency and Advanced Research and Development Activity under Army Research Office contract DAAD-19-02-1-0035, and by the National Science Foundation, grants DMR-0121146 and ECS-0102500.

**References.**

1. K. Hess, *Monte Carlo Device Simulation: Full Band and Beyond* (Kluwer, 1991).
2. C. Moglestue, *Monte Carlo Simulation of Semiconductor Devices* (Chapman & Hall, 1993).
3. C. Jacoboni and L. Reggiani, *Rev. Mod. Phys.* **55**, 645 (1983).
4. S. Das Sarma, J. Fabian, X. Hu, I. Zutic, IEEE Trans. Magn. **36**, 2821 (2000).
5. S. A. Wolf, D. D. Awschalom, R. A. Buhrman, J. M. Daughton, S. von Molnar, M. L. Roukes, A. Y. Chtchelkanova, D. M. Treger, Science **294**, 1488 (2001).
6. S. Das Sarma, Am. Sci. **89**, 516 (2001).
7. D. D. Awschalom, M. E. Flatte, N. Samarth, Sci. Am. **286**, 66 (2002).
8. S. Datta, B. Das, Appl. Phys. Lett. **56**, 665 (1990).
9. B. E. Kane, L. N. Pfeiffer, K. W. West, Phys. Rev. B **46**, 7264 (1992).
10. M. Johnson, Science **260**, 320 (1993).
11. M. E. Flatte, G. Vignale, Appl. Phys. Lett. **78**, 1273 (2001).
12. I. Zutic, J. Fabian, S. Das Sarma, Appl. Phys. Lett. **79**, 1558 (2001).
13. C. Ciuti, J. P. McGuire, L. J. Sham, Appl. Phys. Lett. **81**, 4781 (2002).
14. T. Koga, J. Nitta, H. Takayanagi, Phys. Rev. Lett. **88**, 126601 (2002).
15. X. F. Wang, P. Vasilopoulos, F. M. Peeters, Phys. Rev. B **65**, 165217 (2002).
16. R. G. Mani, W. B. Johnson, V. Narayanamurti, V. Privman, Y.-H. Zhang, Physica E **12**, 152 (2002).
17. J. Schliemann, J. C. Egues, D. Loss, preprint cond-mat/0211678 at www.arxiv.org (2002).
18. R. Fiederling, M. Keim, G. Reuscher, W. Ossau, G. Schmidt, A. Waag, L. W. Molenkamp, Nature **402**, 787 (1999).
19. A. T. Hanbicki, B. T. Jonker, G. Itskos, G. Kioseoglou, A. Petrou, Appl. Phys. Lett. **80**, 1240 (2002).
20. Y. Ohno, R. Terauchi, T. Adachi, F. Matsukura, H. Ohno, Phys. Rev. Lett. **83**, 4196 (1999).
21. F. Mireles, G. Kirczenow, Phys. Rev. B **64**, 024426 (2001).
22. Th. Schapers, J. Nitta, H. B. Heersche, H. Takayanagi, Physica E **13**, 564 (2002).
23. J. Fabian, I. Zutic, S. Das Sarma, Phys. Rev. B **66**, 165301 (2002).
24. G. Schmidt, L. W. Molenkamp, Semicond. Sci. Tech. **17**, 310 (2002).
25. Z. G. Yu, M. E. Flatte, Phys. Rev. B **66**, 235302 (2002).
26. J. Inoue, G. E. W. Bauer, L. W. Molenkamp, preprint cond-mat/0211153 at www.arxiv.org (2002).
27. G. Schmidt, C. Gould, P. Grabs, A. M. Lunde, G. Richter, A. Slobodskyy, L. W. Molenkamp, preprint cond-mat/0206347 at www.arxiv.org (2002).

28. T. Valet, A. Fert, Phys. Rev. B **48**, 7099 (1993).
29. M. Q. Weng, M. W. Wu, J. Appl. Phys. 93, 410 (2003).
30. Y. Qi, S. Zhang, preprint cond-mat/0211674 at www.arxiv.org (2002).
31. D. Hagele, M. Oestreich, W. W. Ruhle, N. Nestle, K. Eberl, Appl. Phys. Lett. **73**, 1580 (1998).
32. H. Sanada, I. Arata, Y. Ohno, Z. Chen, K. Kayanuma, Y. Oka, F. Matsukura, H. Ohno, Appl. Phys. Lett. **81**, 2788 (2002).
33. A. Bournel, V. Delmouly, P. Dollfus, G. Tremblay, P. Hesto, Physica E **12**, 86 (2001).
34. A.A. Kiselev, K.W. Kim, Phys. Rev. B **61,** 13115 (2000).
35. S. Saikin, M. Shen, M.-C. Cheng, V. Privman, preprint cond-mat/0212610 at www.arxiv.org (2002).
36. V. V. Mitin, V. A. Kochelap, M. A. Stroscio, *Quantum Heterostructures. Microelectronics and Optoelectronics* (Cambridge University Press, 1999).
37. G. Fishman, G. Lampel, Phys. Rev. B **16**, 820 (1977).
38. K. Blum, *Density Matrix Theory and Applications* (Plenum Press, 1996).
39. G. Dresselhaus, Phys. Rev. **100**, 580 (1955).
40. M. I. Dyakonov, V. Yu. Kachorovskii, Sov. Phys. Semicond. **20**, 110 (1986).
41. Yu. Bychkov, E. I. Rashba, J. Phys. C **17**, 6039 (1984).
42. M. I. D'yakonov, V. I. Perel', JETP **33**, 1053 (1971).
43. R. J. Elliott, Phys. Rev. **96**, 266 (1954).
44. M. Cheng and E. E. Kunhardt, *J. Appl. Phys.* **63**, 2322 (1988).
45. K. Tomizawa, *Numerical Simulation of Submicron Semiconductor Devices* (Artech House, 1993).
46. L. Reggiani, *Hot Electron Transport* (Springer-Verlag, 1985).
47. J. Nitta, T. Akazaki, H. Takayanagi, Phys. Rev. Lett. **78**, 1335 (1997).
48. M. Cardona, N. E. Christensen, G. Fasol, Phys. Rev. B **38**, 1806 (1988).
49. M. V. Fischetti, S. E. Laux, *Damocles Theoretical Manual* (IBM Corporation, April 1995).
50. V. F. Gantmakher, Y. B. Levinson, *Carrier Scattering in Metals and Semiconductors* (Elsevier, 1987).
51. M. M. Glazov, E. L. Ivchenko, preprint cond-mat/0301519 at www.arxiv.org (2003).
52. M. Cahay, S. Bandyopadhyay, preprint cond-mat/0301052 at www.arxiv.org (2003).